\documentclass[prd,aps,nofootinbib,showpacs,showkeys,preprintnumbers,
axodraw]{revtex4}
\usepackage{graphicx,epsf,amsfonts,amssymb,amsbsy}
\newcommand{\vev}[1]{\langle#1\rangle}
\newcommand{\mat}{\left ( \begin{array}}
\newcommand{\emat}{\end{array} \right )}
\newcommand{\vect}{\left ( \begin{array}{c}}
\newcommand{\evect}{\end{array} \right )}

\preprint{HU-EP-05/27}

\begin{document}

\title{ \bf Pion condensation in quark matter with finite
baryon density}
\author{D.~Ebert}
\email{debert@physik.hu-berlin.de}
\affiliation{Institut f\"ur Physik,
Humboldt-Universit\"at zu Berlin, 10115 Berlin, Germany}
\author{K.\,G.~Klimenko}
\email{kklim@ihep.ru}
\affiliation{Institut f\"ur Physik,
Humboldt-Universit\"at zu Berlin, 10115 Berlin, Germany  and
\\Institute
of High Energy Physics, 142281, Protvino, Moscow Region, Russia}

\begin{abstract}
The phase structure of the Nambu -- Jona-Lasinio model at zero
temperature and in the presence of baryon- and isospin chemical
potentials is investigated. It is shown that in the chiral limit and
for a wide range of model parameters there exist two different phases
with pion condensation. In the first, ordinary phase, the baryon
density is zero and quarks are gapped particles. In the second,
gapless pion condensation phase, there is no energy cost for
creating only u-or both u and d quarks, and the density of baryons is
nonzero.
\end{abstract}


\maketitle

According to a well-known point of view \cite{migdal}, pionic degrees
of freedom and, especially,
the pion condensation
phenomenon might play a significant role in the description of
different nuclear matter effects,
heavy-ion collisions, and the physics of compact stars.
In reality, dense baryonic
matter obeys an isospin asymmetry, i.e. the densities of up- and down
quarks are different.
In order that QCD adequately describes the isotopically asymmetric
matter, usually the isospin chemical
potential $\mu_I$ is introduced into the theory. It is well-known
that in QCD with finite $\mu_I$
a nonzero pion condensate is generated if $\mu_I$
is greater than the pion mass $m_\pi$. This result was obtained in
the framework of an effective chiral Lagrangians approach \cite{son}
as well as in QCD lattice calculations,
performed at zero or small values of the baryon chemical potential
$\mu_B$ \cite{kogut}. However, these two approaches are not
applicable for the description of an isotopically
asymmetric matter at moderate baryon density.
To overcome this problem, it was recently proposed to study the QCD
phase diagram on the basis of
Nambu -- Jona-Lasinio (NJL)-type models, which contain quarks as
microscopic degrees of freedom, in the presence of a baryon chemical
potential $\mu_B$ and an isospin $\mu_I$ one. In this way the
influence of $\mu_B$, $\mu_I$ on both the chiral symmetry restoration
effect \cite{toublan} and the formation of
color superconducting (CSC) dense baryonic matter \cite{toki} was
considered, but with a zero pion condensate (for earlier
investigations of the NJL phase diagram in the plane of
temperature-baryon density, but without isospin asymmetry, i.e.
$\mu_I=0$, see, e.g., \cite{zahed}).
In addition, it was found that if the
influence of $\mu_I$ is combined with the condition of electric
charge neutrality, then essential changes of
quasiparticle dispersion relations might occur in CSC matter.
As a consequence, it is transformed into a so-called gapless CSC
matter, since new gapless quasiparticle degrees of freedom appear
(see, e.g., reviews \cite{hs}). Recently, the pionic
condensation effect was investigated in some NJL models
at nonzero values of $\mu_B$ and $\mu_I$ \cite{barducci,zhuang}.
However, in these papers the existence of a homogeneous dense
baryonic phase with pion condensation was not investigated. 
So, the question whether it is possible to describe such a phase
in the framework of NJL models is opened up to now.

To answer this question, we study in present paper the phase
structure of the NJL model in terms of $\mu_B$ and $\mu_I$. 
We show that, depending on the parameters of the
chosen NJL model, a homogeneous phase of an isospin asymmetric quark
matter with pion condensation and with finite baryon density may
appear. Since quarks are gapless in this phase, we call it gapless
pion condensed (GPC) phase. In addition, a pion condensed phase with
zero baryon charge is also presented in the phase structure of the
model. In contrast to GPC phase, quarks are gapped in it. In the
following we restrict ourselves to values $\mu_B\le 1$ GeV in order
to avoid the inclusion of a Cooper pairing of quarks into
consideration. (At greater $\mu_B$ the diquark condensation should be
taken into account \cite{1,2}.)

Let us study the properties of quark matter at moderate baryon and
isospin densities and at zero
temperature in the framework
of the NJL model with two quark flavors
\begin{eqnarray}
&&  L_q=\bar q\Big [\gamma^\nu i\partial_\nu-
m+\mu\gamma^0+\delta\mu\tau_3\gamma^0\Big ]q+ G\Big [(\bar qq)^2+
(\bar qi\gamma^5\vec\tau q)^2\Big ],
  \label{1}
\end{eqnarray}
where $\mu\equiv\mu_B/3$ is the quark number chemical potential and
the quantity $\delta\mu$ is a
half of the isospin chemical
potential, i.e. $\delta\mu\equiv \mu_I/2$. In (\ref{1}) $\tau_i$
($i=1,2,3$) are Pauli matrices.
Evidently, at nonvanishing current quark masses, $m\ne 0$, the
Lagrangian (\ref{1}) is invariant
under the baryon
$\rm U_B(1)$ symmetry and the parity transformation P. Moreover,
without the $\mu_I$-term it
is also invariant under the isospin $\rm SU_I(2)$ group which is
reduced to
$\rm U_{I_3}(1)$ at $\mu_I\ne 0$.
Usually, the properties of the ground state of the model (\ref{1})
are characterized by two order parameters.
The first one is the quark condensate $\vev{\bar qq}$ which is
responsible for the chiral symmetry
breaking and does not spoil the parity and isotopical symmetry.
 The second one, called pion condensate,
is the ground state expectation value of the form $\vev{\bar
q\gamma^5\tau_1 q}$. If it is nonzero,
spontaneous breaking of parity and isotopic symmetry
occur in the system. \footnote{One may, in addition, suppose that
$\langle\bar q\gamma^5\tau_2 q\rangle$ is also nonzero. Then, using
the $\rm U_{I_3}(1)$-rotations,
this quantity can be reduced to zero.} At $\mu_I =0$ only the quark
condensate is relevant to the physics
of the model, but at nonzero $\mu_I$ it is desirable to take into
account both condensates.
In the last case the competition between them is governed by the
thermodynamic potential (TDP) which in the mean field approximation
has the following form
which can be obtained, using, e.g., the technique of \cite{3}:
\begin{eqnarray}
\Omega(M,\Delta)&=&\frac{(M-m)^2+\Delta^2}{4G}-3\sum_a\int\frac{d^3p}
{(2\pi)^3}~|E_a|=\nonumber\\
&=&\frac{(M-m)^2+\Delta^2}{4G}-6\int\frac{d^3p}{(2\pi)^3}\Big\{E_
\Delta^-+E_\Delta^++(\mu-E_\Delta^-)\theta(\mu-E_\Delta^-)+
(\mu-E_\Delta^+)\theta(\mu-E_\Delta^+) \Big\}.
\label{2}
\end{eqnarray}
The summation in the first line of (\ref{2}) runs over all
quasiparticles ($a=u,d,\bar u,\bar d$), where
\begin{eqnarray}
E_u=E_\Delta^--\mu,~~~~~~&& E_{\bar u}=E_\Delta^++\mu,\nonumber\\
E_d=E_\Delta^+-\mu,~~~~~~&& E_{\bar d}=E_\Delta^-+\mu,
\label{3}
\end{eqnarray}
\begin{figure}
  \centering
  \includegraphics[width=10cm]{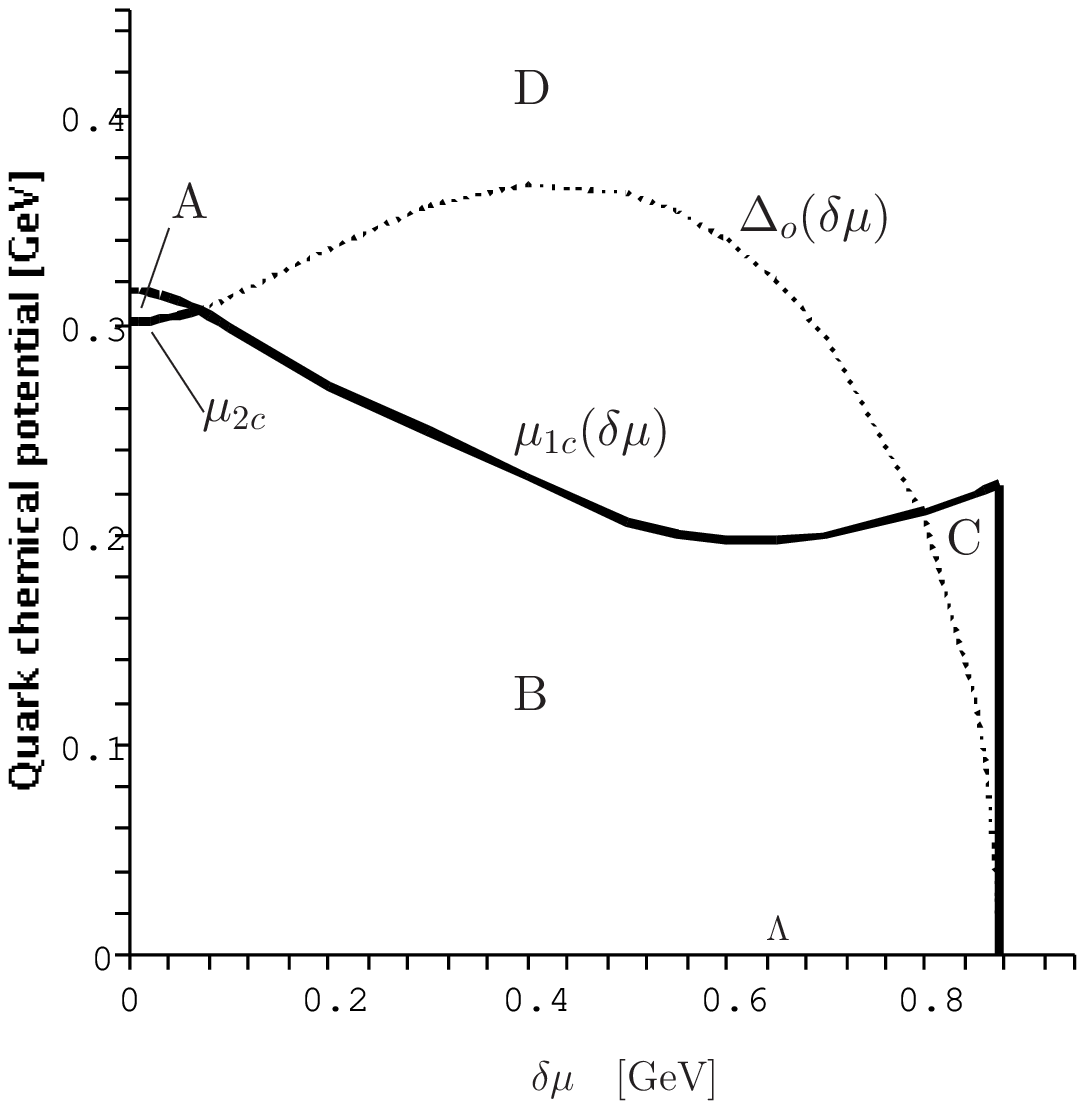}
\caption{The phase portrait of the model at $m=0$ in terms of $\mu$,
$\delta\mu$ for the set 1 of   parameters, $\Lambda =0.65$ GeV,
$G=5.01$ GeV$^{-2}$.  The region D corresponds to the fully symmetric
phase with $M=0$, $\Delta=0$. A, B, and C are phases with pion
condensation. The solid lines $\mu_{1c}(\delta\mu)$ and $\mu_{2c}
(\delta\mu)$ are the critical curves of first order phase
transitions. The dotted line, $\Delta_o(\delta\mu)$, is the pion
condensate at $\mu=0$. The baryon density is zero in the phase B and
nonzero in A, C, D. There are gapless quarks in the phases A, C, D.
In the phase B quarks are gapped.} \label{plot:1}
\end{figure}
and $E_\Delta^\pm=\sqrt{(E^\pm)^2+\Delta^2}$, $E^\pm=E\pm\delta\mu$
and $E=\sqrt{\vec p^2+M^2}$. The factor 3 in front of the summation
symbol in (\ref{2}) indicates the three-fold degeneracy of each
quasiparticle in color. Moreover, in order to avoid
usual ultraviolet divergences, the integration region in
(\ref{2}) is restricted by a cutoff $\Lambda$, i.e. $|\vec
p|<\Lambda$, where $\Lambda \le 1$ GeV. The gap coordinates of the
global minimum point of the function $\Omega(M,\Delta)$ are connected
with condensates in the following way:
\begin{eqnarray}
M=m-2G\vev{\bar qq},~~~~~~\Delta=-2Gi\vev{\bar q\gamma^5\tau_1 q},
\label{31}
\end{eqnarray}
so if $\Delta$ is nonzero in the global minimum point (GMP), then
parity as well as isotopic symmetry are spontaneously broken and the
pion condensation phase is realized.
Note that the quark gap $M$ is just the dynamical quark mass. From
(\ref{2}) it is possible to obtain
the gap equations
\begin{eqnarray}
0=\frac{\partial\Omega (M,\Delta)}{\partial M}&\equiv&
\frac{M-m}{2G}-6M\int\frac{d^3p}{(2\pi)^3E}\Big\{\frac{\theta(E_
\Delta^+-\mu)E^+}{E_\Delta^+}+
\frac{\theta(E_\Delta^--\mu)E^-}{E_\Delta^-} \Big\},\nonumber\\
0=\frac{\partial\Omega (M,\Delta)}{\partial\Delta}&\equiv&
\frac{\Delta}{2G}-6\Delta\int\frac{d^3p}{(2\pi)^3}\Big\{\frac{\theta(
E_\Delta^+-\mu)}{E_\Delta^+}+
\frac{\theta(E_\Delta^--\mu)}{E_\Delta^-} \Big\}.
\label{4}
\end{eqnarray}
\begin{figure}
  \centering
  \includegraphics[width=10cm]{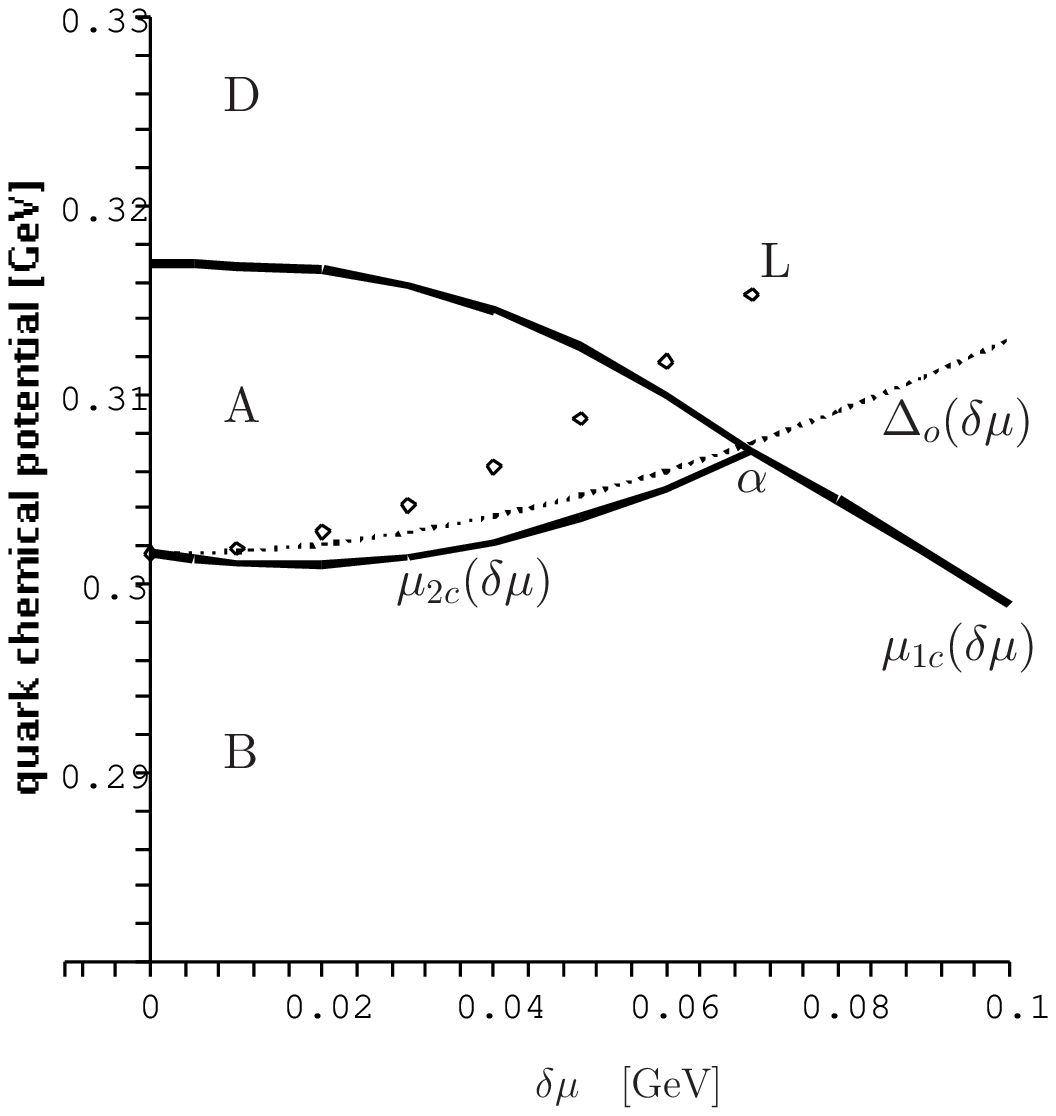}
  \caption{The gapless phase A of the NJL model at $m=0$. The
  notations are the same as in Fig. 1.
  The schematic curve L divides A into the region with only $u$
  gapless quarks
(to the right of L) and the region with both $u$ and $d$ gapless
quarks (to the left of L).
In the triple point $\alpha$ three phases A, B, D coexist.}
\label{plot:2}
\end{figure}

We will investigate the phase structure of the NJL model (\ref{1})
both at $\mu_I \ne 0$ and $\mu\ne 0$. 
If the current quark mass $m$ is not zero, then the pion condensation
in the framework of NJL approach is occured at $\mu_I>m_\pi$
\cite{barducci,zhuang}. Just for these values of $\mu_I$ it is
reasonable to search for the pion condensed phase with finite baryon
density. In general, it is a rather hard task, so in the present
paper we restrict ourselves by the zero current quark mass case only
($m=0$, $m_\pi=0$), thus studying the phase structure of the model
for all $\mu_I>0=m_\pi$. 
At $m=0$ and $\mu_I \ne 0$ the symmetry group of the initial model is
$\rm U_B(1)\times U_{I_3L}(1)\times U_{I_3R}(1)\times P$. The system
of the gap equations (\ref{4}) has three different solutions: {\bf
i)} $M=0$, $\Delta =0$, {\bf ii)} $M\ne 0$, $\Delta =0$, {\bf iii)}
$M=0$, $\Delta\ne 0$. Having all the solutions of the gap equations,
it is necessary to compare the values of the TDP on them, selecting
by this way the minimal one, i.e. the GMP.
If the GMP of the TDP is achieved on the solution {\bf i)}, then the
ground state of our system is invariant under the initial symmetry
group of the model. The solution of the type {\bf ii)} corresponds to
the phase, symmetrical under $\rm U_B(1)\times U_{I_3}(1)\times P$
group, where $U_{I_3}(1)$ acts in the following way: $q\to\exp
(i\alpha\tau_3) q$.
Finally, if the GMP of the TDP lies in the point of the form {\bf
iii)}, then in the ground state the quark condensate is zero, but the
pion condensate is nonzero. In this case the pion condensation phase
is realized in the model, and the ground state is symmetrical under
$\rm U_B(1)\times U_{AI_3}(1)$, which transforms left- and right
quark flavor doublets in the following way: $q_L\to\exp
(i\alpha\tau_3) q_L$, $q_R\to\exp (-i\alpha\tau_3) q_R$.
Generally, at $m=0$, due to a symmetry of the TDP, it is sufficient 
to study its GMP only in the region $M,\Delta\ge 0$ and for
nonnegative values of $\delta\mu$, $\mu$. For the moment let us fix
the parameters as $G=5.01$ GeV$^{-2}$, $\Lambda =0.65$ GeV (set 1).
The dependence of quark matter properties on the different sets of
model parameters $G,\Lambda$ will be discussed later.

Following the way, described above, we obtain for set 1 of parameters
the model phase structure that is represented in Fig. 1 in terms of
$\delta\mu=\mu_I/2>0$ and $\mu$. There, the region D corresponds to
the totally symmetric phase of the model, in which $M=0$
and $\Delta=0$. This phase is separated from the regions A, B, C by
the critical curve $\mu=\mu_{1c}(\delta\mu)$ of first order phase
transitions. In A, B, and C
the GMP of the TDP has the form {\bf iii)}, so these
regions correspond to phases with pion condensation. 

In the phase B the pion condensate depends only on $\delta\mu$ (not
on $\mu$), so
$\Delta\equiv\Delta_o(\delta\mu)$, where $\Delta_o(\delta\mu)$ is the
pion condensate at $\mu=0$.
Since for all points of the region B we have the relation
$\mu<\Delta_o(\delta\mu)$ (see Fig. 1,
where $\Delta_o(\delta\mu)$ is depicted by the dotted line),
it is clear that the value of the TDP in the corresponding GMP point,
$\Omega_{\rm min}$, does
not depend on the quark chemical potential $\mu$
(this follows from the second line of (\ref{2})).
So, in the phase B the baryon density $n_B$ is identically zero
($n_B=-\partial\Omega_{\rm min}/
(3\partial\mu)$).
In contrast, in the phase A, which is presented in more details in
Fig. 2, the pion condensate depends both on $\mu$ and $\delta\mu$,
i.e. $\Delta\equiv\Delta(\mu,\delta\mu)$.
Hence in the phase A the baryon density is nonzero. Since on the
curve $\mu=\mu_{2c}(\delta\mu)$ of Figs 1,2 we have
$\Delta(\mu,\delta\mu)\ne \Delta_o(\delta\mu)$,
the transition from the region A to B is a discontinuous first order
phase
transition. In the so-called triple point $\alpha$ of Fig. 2 three
phases A, B, D coexist.
In the phase C that is separated from B by the dotted line
$\mu=\Delta_o(\delta\mu)$, the pion condensate
also does not depend on $\mu$. Moreover, here it is the same
as in the phase B, i.e. $\Delta\equiv\Delta_o(\delta\mu)$. However,
since for the points of the region
C we have $\mu>\Delta_o(\delta\mu)$ (see Fig. 1), the minimal value
$\Omega_{\rm min}$ of the TDP (\ref{2}) 
is changed with $\mu$. As a result, the baryon
density in the phase C is nonzero as
 well. Note that the phase C is located at values
 $\delta\mu>\Lambda$.
In this case one might believe that it is an artefact of the NJL
model and not adequate to the real
situation. On the other hand, the phase A corresponds to rather small
values
of $\mu$ and $\mu_I$, so it might be relevant to real physics and
needs some more discussions.

\begin{figure}
  \centering
  \includegraphics[width=10cm]{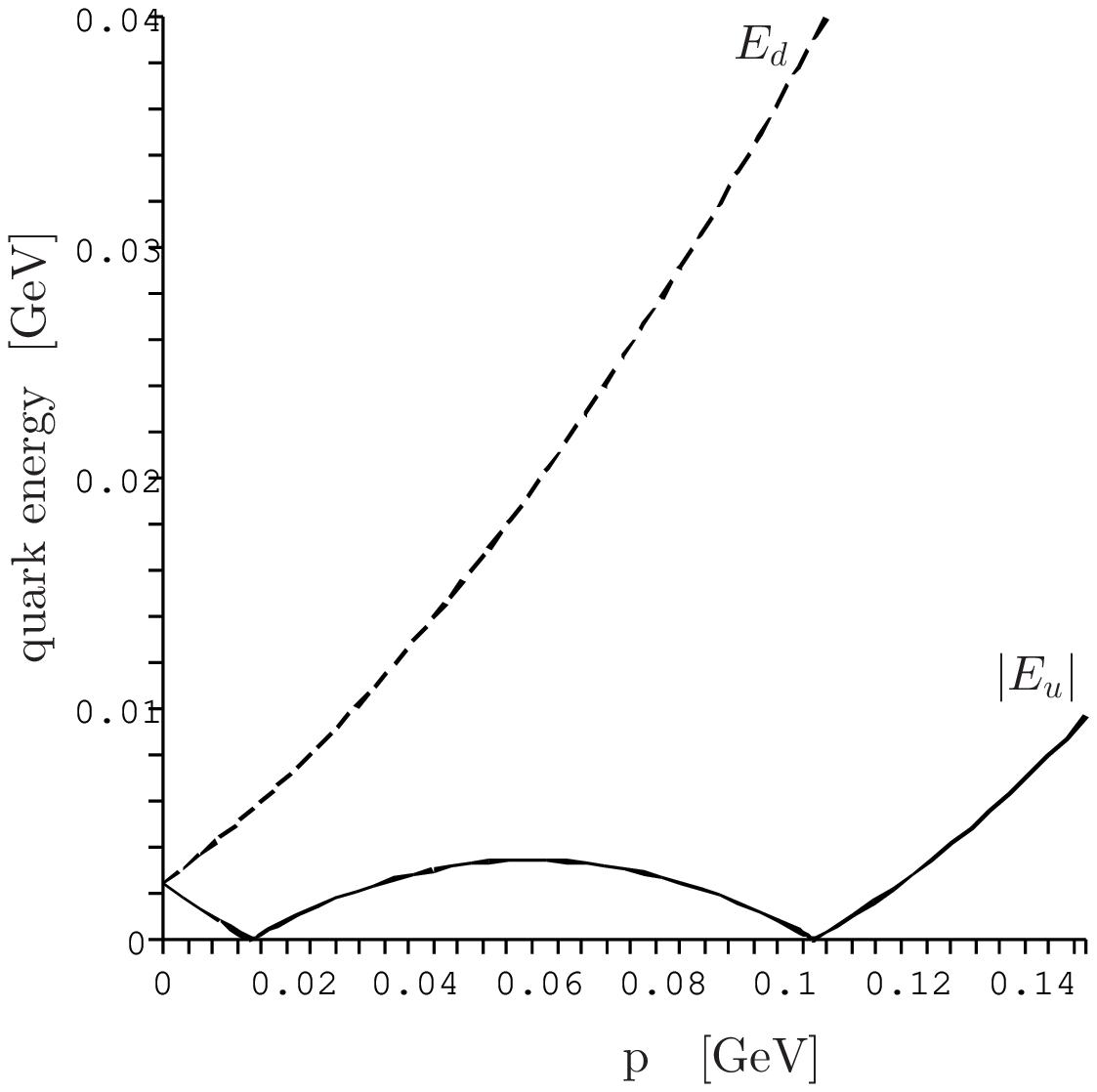}
  \caption{Typical behaviour of quark energies in the phase A and to
  the right of L (see Fig. 2).
  The case presented corresponds to $\delta\mu=0.06$ GeV,
$\mu\approx 0.306$ GeV. In this case the pion condensate takes the
value
$\Delta(\mu,\delta\mu)\approx 0.303$ GeV.}
\label{plot:3}
\end{figure}

As it is easily seen from Fig. 2, for the set 1 of $G$ and $\Lambda$
the height of the region A, i.e. the
quantity $H=\mu_{1c}(0)-\Delta_o(0)$ (in the model under
consideration, at $m=0$ the value $\Delta_o(0)$
is nothing else than the dynamical (constituent)
quark mass $M_q\approx 0.3$ GeV at $\mu =0$, $\delta\mu =0$), is
approximately 15 MeV, whereas the
extention of it along the $\delta\mu$-axis reaches 70 MeV (for $\mu_I
=140$ MeV).
Apart from the baryon density, there is another
physical characteristic which is different in A and B phases. It is
just the minimal energy,
necessary for creating a single quark in the ground state.
Since the quark condensate is equal
to zero both in A and B phases, one should take $M=0$ in formulae
(\ref{3}) in order to obtain
the energies of $u$- and $d$ quarks in the pion condensed phases A
and B. So,
\begin{eqnarray}
E_u=\sqrt{(p-\delta\mu)^2+\Delta^2}-\mu,~~~~~~E_d=\sqrt{(p+\delta\mu)
^2+\Delta^2}-\mu,
\label{5}
\end{eqnarray}
\begin{figure}
  \centering
  \includegraphics[width=10cm]{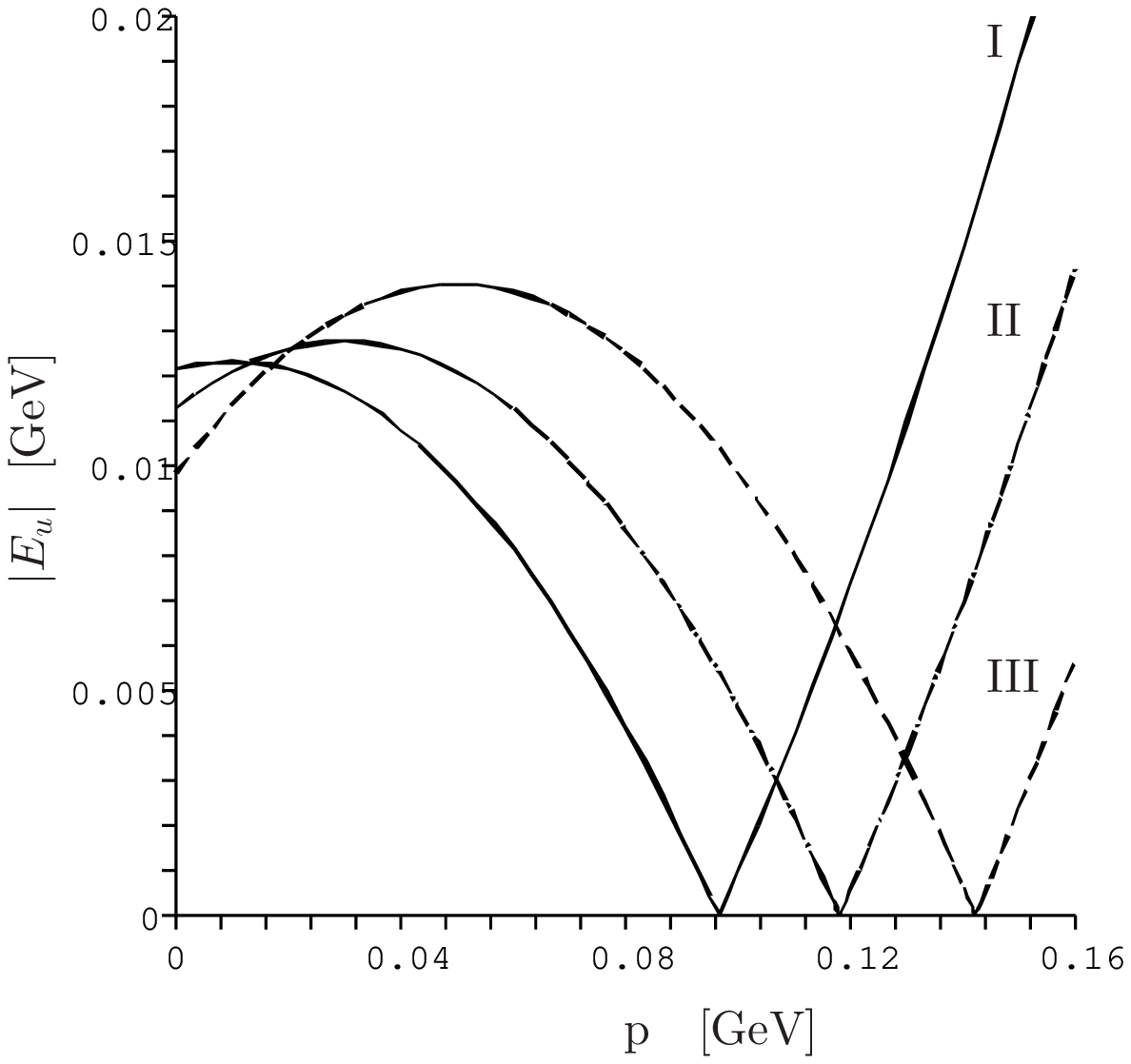}
  \caption{Energies of $u$ quarks vs three momentum $p$ for three
  different points of the phase A that are to the left of the curve L
  (see Fig. 2). I - the case of
$\delta\mu=0.01$ GeV,
$\mu\approx 0.31$ GeV, $\Delta(\mu,\delta\mu)\approx 0.298$ GeV. II -
the case of $\delta\mu=0.03$ GeV,
$\mu\approx 0.31$ GeV, $\Delta(\mu,\delta\mu)\approx 0.297$ GeV. III
- the case of $\delta\mu=0.05$ GeV,
$\mu\approx 0.31$ GeV, $\Delta(\mu,\delta\mu)\approx 0.296$ GeV.}
\label{plot:4}
\end{figure}
where $p=|\vec p|$. If $\Delta >\mu$, the minimal values of the
quantities $E_{u,d}$ are greater than zero.
This fact means that for creating a single quark
in the phase B, where $\Delta\equiv\Delta_o(\delta\mu) >\mu$, one
needs an energy that is greater
than some finite (nonzero) amount (also
called gap). Hence, quarks are gapped in the B phase. In contrast,
for all points
of the phase A the relation $\Delta\equiv\Delta (\mu,\delta\mu)<\mu$
is fulfilled. Hence, in this phase
the minimal energy necessary for creating $u$- or both $u$- and $d$
quarks is always zero.
So, in the phase A we have gapless quarks.
The line L that is defined implicitly by the equation
$\mu=\sqrt{(\delta\mu)^2+\Delta^2 (\mu,\delta\mu)}$
 divides the region A into two parts (in Fig. 2 the line L
is represented very schematically).
To the right of L only $u$ quarks are gapless in the phase A. Typical
behaviours of quark energies vs $p$ in this region are shown in Fig.
3.
(Note, $|E_{u}|$ turns into zero at two different
values of $p$.) To the left of L both $u$- and $d$ quarks are gapless
ones. This situation is illustrated in
Figs 4 and 5, where for $\mu=0.31$ GeV the energy of the $u$ quark is
depicted for three different values of $\delta\mu$ (Fig. 4), whereas
in Fig. 5 the energy of the $d$
 quark is represented for the same values of $\mu$ and $\delta\mu$.
 It is
easily seen that to the left of L the energies of $u$- and $d$ quarks
turn into zero only
in a single point of the $p$ variable.
Due to these features, it is naturally to call the phase A the
gapless pion condensed (GPC) phase. (Note, in the phases C and D
there are also gapless quasiparticle
excitations.)

\begin{figure}
  \centering
  \includegraphics[width=9cm]{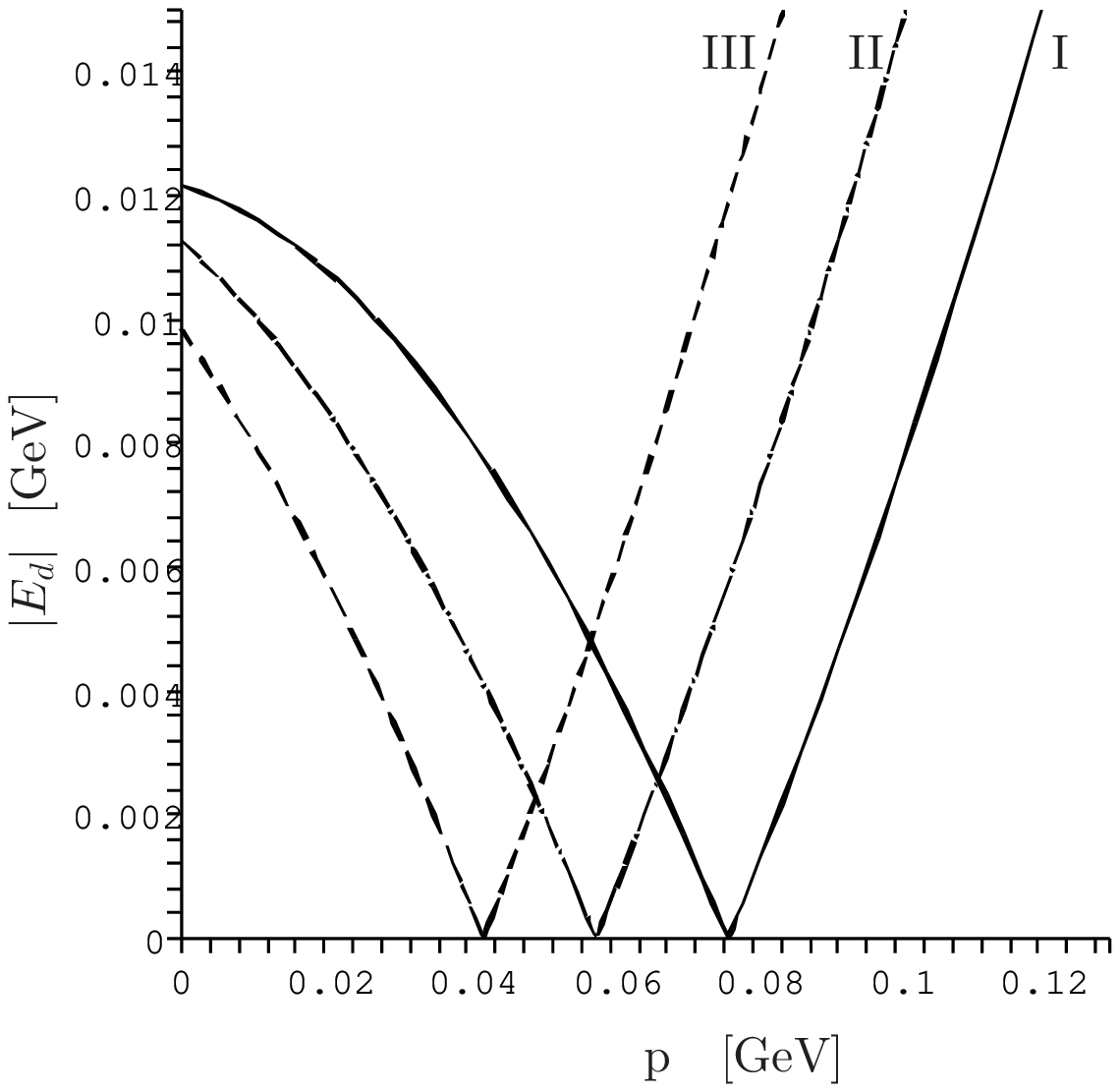}
\caption{The analogue curves as in the Fig. 4, but for the $d$
quarks.}
\label{plot:5}
\end{figure}

Up to now in considering the GPC phase we have dealed only with the
set 1 of parameters $G$ and $\Lambda$.
For this set the constituent quark mass $M_q$ in the vacuum, i.e.
at zero $\mu$ and $\mu_I$, is approximately 0.3 GeV. It follows from
the relation, connecting $G$,
$\Lambda$, and $M_q$ \cite{ek}:
\begin{eqnarray}
\frac{\pi^2}{6G}=\Lambda\sqrt{\Lambda^2+M_q^2}-M_q^2\ln \Big
(\frac{\Lambda+\sqrt{\Lambda^2+M_q^2}}{M_q} \Big ).
\label{6}
\end{eqnarray}
Using (\ref{6}), it is possible to study the dependence of the height
$H=\mu_{1c}(0)-\Delta_o(0)$
of the GPC phase A on the parameters $\Lambda$ and $M_q$
(see Fig. 6). Clealy, positive values of $H$ indicates on the
presence of GPC phase. In this figure
one can see a set of curves. Each curve represents the behaviour of
$H$ vs $\Lambda$ at fixed $M_q$.
(The result $H=15$ MeV, obtained above for the set 1,
corresponds to the curve with $M_q=0.3$ GeV at $\Lambda =0.65$ GeV.)
It follows from Fig. 6 that the GPC
phase is presented in the phase structure of the NJL model
for a rather wide range of model parameters. The smaller $M_q$, the
greater the height of the GPC phase
at fixed $\Lambda$.
In this context, it is worth noting that some parametrizations of the
NJL model correspond to
rather small values of $M_q$. For example,
in \cite{ev} the constituent quark mass was taken as small as 0.233
GeV, whereas $\Lambda =1$ GeV.
Hence, the height of the GPC phase in this parametrization reaches
the value 50 MeV. In contrast, in the paper \cite{barducci} the
quantities $M_q=0.428$ GeV and
$\Lambda =0.58$ GeV were used for the consideration of the NJL model
at nonzero $\mu_B$ amd $\mu_I$. It is clear from Fig. 6 that for such
values of parameters the
GPC phase is absent in the phase structure, so it is no wonder
why in \cite{barducci} the homogeneous phase, describing quark matter
with finite baryon density
and pion condensation, was not found.
\vspace{0.5cm}

In summary, we have investigated the phase structure of the two
flavor NJL model (\ref{1}) at finite
baryon and isospin chemical potentials for a current quark mass $m$
taken to be zero.
It was shown that at the cutoff value $\Lambda=0.65$ GeV and for the
constituent
quark mass $M_q=0.3$ GeV (or $G=5.01$ GeV$^{-2}$) there are three
different phases in the model, A, B, D, at $0<\delta\mu <\Lambda$.
The phase D corresponds to quark
matter with finite baryon density and zero pion condensation.
In the phases A and B pions are condensed. However, baryon density in
the phase B is zero, but in A
it is nonzero. Moreover, the phase B is a gapped one, since creation
of a single
quark in the ground state of this phase necessarily costs some finite
amount of energy. In
contrast, in the phase A $u$-quarks or both $u$- and $d$-quarks are
gapless. These three phases can
coexist in the triple point $\alpha$ (see Fig. 2), since transitions
from one phase to another are
discontinuous. 

It is well-known that pion condensation takes place if $\mu_I$ is
greater than the pion mass $m_\pi$ (see, e.g., \cite{zhuang}). For
simplicity, in the present paper we have considered the zero current
quark mass case, where pion mass is zero. That is why the GPC phase
is observed at rather small values of $\mu_I$. In a more realistic
situation with $m\ne 0$ the pion mass equals approximately 140 MeV.
We believe that in this case GPC phase might be predicted as well,
but at $\mu_I>m_\pi$.
\begin{figure}
  \centering
  \includegraphics[width=9cm]{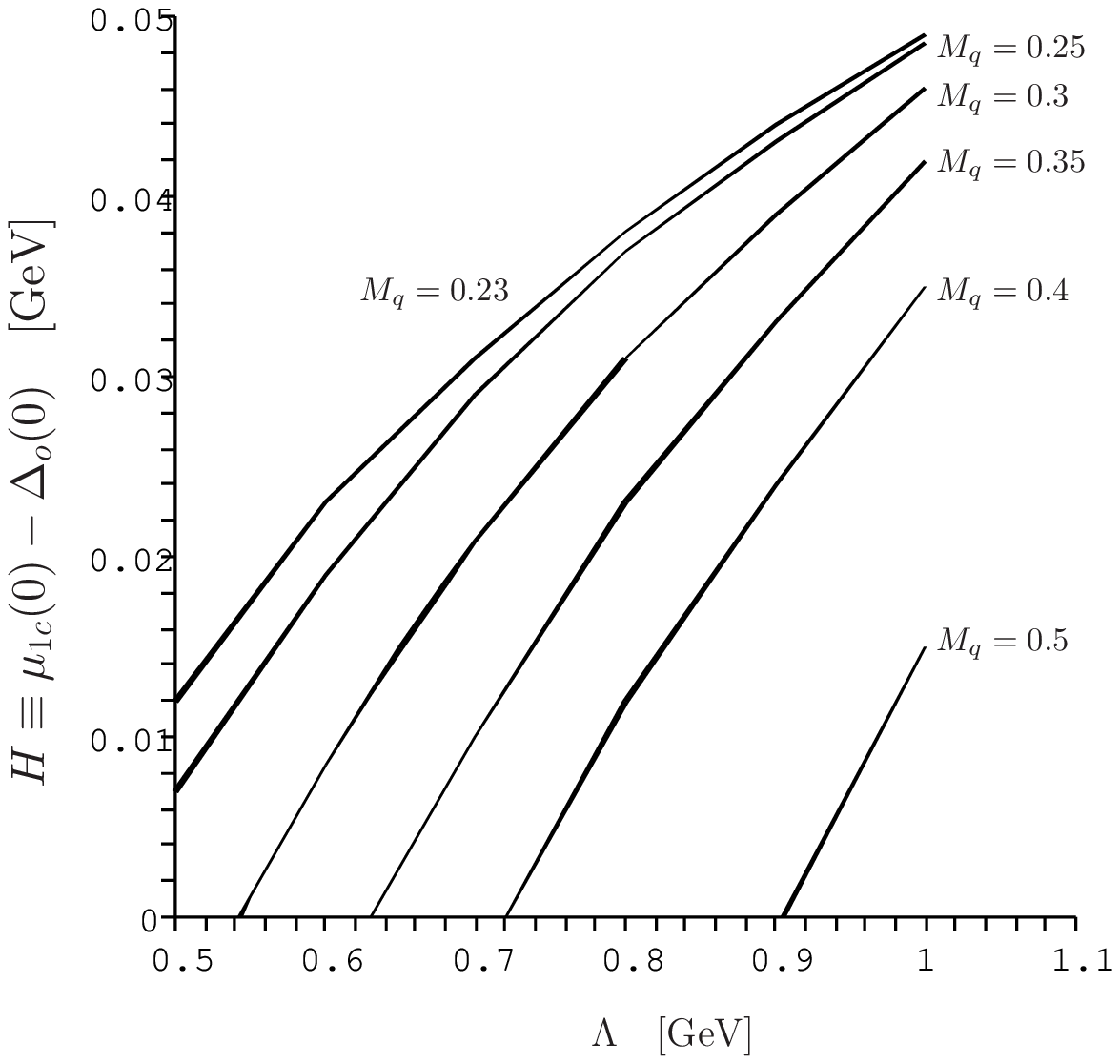}
\caption{The heights $H$ of the gapless phase A vs cutoff $\Lambda$
for different fixed values of constituent quark masses $M_q$ (in
GeV).}
\label{plot:6}
\end{figure}

In conclusion, there arises the interesting question,
whether a pion condensate
with finite baryon number density might exist in nature.
\vspace{0.5cm}

{\bf Acknowledgments:} One of us (K.G.K.) thanks Prof. M.
Mueller-Preussker and the colleagues of the
Institute of Physics of the Humboldt University (Berlin) for kind
hospitality.
This work was supported in part by DFG-project 436 RUS 113/477/0-2
and RFBR grant No. 05-02-16699.



\end{document}